# Statistical mechanics characterization of spatio-compositional inhomogeneity

## Ryszard Piasecki[*]

*Faculty of Chemistry, University of Opole, Oleska 48, 45-052 Opole, Poland*

---

**Abstract**

On the basis of a model system of pillars built of unit cubes, a two-component entropic measure for the multiscale analysis of spatio-compositional inhomogeneity is proposed. It quantifies the statistical dissimilarity per cell of the actual configurational macrostate and the theoretical reference one that maximizes entropy. Two kinds of disorder compete: i) the spatial one connected with possible positions of pillars inside a cell (the first component of the measure), ii) the compositional one linked to compositions of each local sum of their integer heights into a number of pillars occupying the cell (the second component). As both the number of pillars and sum of their heights are conserved, the upper limit for a pillar height $h_{max}$ occurs. If due to a further constraint there is the more demanding limit $h \leq h^* < h_{max}$, the exact number of restricted compositions can be then obtained only through the generating function. However, at least for systems with exclusively composition degrees of freedom, we show that the neglecting of the $h^*$ is not destructive yet for a nice correlation of the $h^*$-constrained entropic measure and its less demanding counterpart, which is much easier to compute. Given examples illustrate a broad applicability of the measure and its ability to quantify some of the subtleties of a fractional Brownian motion, time evolution of a quasipattern [28,29] and reconstruction of a laser-speckle pattern [2], which are hardly to discern or even missed.

---

## Contents

## 1. Introduction

The limited configurational information for disordered materials can be obtained with a set of lower-order of *n*-point correlation functions [1]. Thus, it is highly nontrivial to predict on this basis an effective macroscopic property. However, the reconstruction of configurations with target pair correlations although not entirely perfect is still possible [2,3]. The inverse statistical-mechanical methods to find interaction potentials, which correspond to optimally stable target structures, for instance, in soft matter systems have been recently reviewed in Ref. [4]. They incorporate structural information that generally accounts for almost complete morphological features. According to opinion of the author of the above review article, such inverse approaches could be applied to create "novel materials with varying degrees of disorder, thus extending the traditional idea of self-assembly to incorporate not only crystals but amorphous and quasicrystal structures" [4]. Reproducing complex multiphase microstructures in the context of predicting their effective physical properties has been also recently discussed in Ref. [5].

---

[*] Tel.: +48 77 4527116; fax: +48 77 4527101.
    *E-mail address*: piaser@uni.opole.pl.

The another approach to the reconstruction of disordered materials might provide multiscale entropic measure (descriptor) of average spatial inhomogeneity for random systems of finite size objects (FSOs) [6,7]. For a binary pattern FSO can be represented by black pixels 'interacting' with each other through mutual exclusion. It should be stressed that within this kind of combinatorial approach we exactly evaluate a possible number of realizations (microstates) for a given configurational macrostate. It seems that its Tsallis version [8] related to FSOs may be also (in future projects) taken into account [9], because different nonextensivity $q$-parameters at various *spatial* scales can be chosen for systems far from equilibrium with fluctuations in temperature or energy dissipation rate [10,11], what makes the Tsallis entropies quasi-additive. Recently, although in a different context, it was pointed out how finite size objects comes into play when the process of gaining information accounts for their size [12]. On the other hand, the notion of FSO has been utilized again to easy quantify of the inhomogeneity of greyscale images. The simple extension of the previous binary measure to a grey level inhomogeneity one has been proposed [13] as well as the general entropic descriptor of a complex behaviour [14]. The both developments are adaptable for advanced methods of reconstruction of systems with a grey level multiscale complex structure.

The present development, that is the innovative two-component entropic measure for multiscale analysis of spatio-compositional inhomogeneity, primarily was thought up for a model system of decomposable pillars built of unit cubes. However, the *non-zero* integer heights of the model pillars can represent *shades of grey* shifted, for instance, from the standard range 0−255 to 1−256 in the case of 8-bit greyscale images. So, in our approach each of unoccupied (by pillars) positions is encoded now with zero in integer valued matrix that represents a system of pillars or equivalently, a greyscale image for a general case (without or with missing pixels). As we see further, our concept differs from those based on Shannon entropy and probability distributions. For example, in Ref. [15] it was assumed that the "homogeneity value of a pixel" at location $(i, j)$ is proportional to the uniformity of the region $3 \times 3$ surrounding the pixel. Then, a normalized homogeneity measurement was defined to enhance the contrast of grey level image. However, since the every square window should be centred at integer position $(i, j)$, this approach supposedly omits square windows of even side. Another measure of quality of mixing achieved at a given scale of observation, so-called colour homogeneity index is calculated by using joint probabilities [16]. To estimate probabilities by using appropriate frequencies the large number of particles is needed what sets an upper limit for the number of bins (a lower scale limit) for characterization of a system.

Instead, free of such limitations, we use the generalized configurational macrostates and Boltzmann entropy to develop a novel tool for multiscale analysis of greyscale images of a broad applicability. Our two-component entropic measure can quantify at different length scales: i) *spatial* inhomogeneity (SpIN) − the first component, and ii) *compositional* inhomogeneity (CoIN) − the second component, for numerous systems, which short exemplary list is given below. If the both components of the proposed measure are non-zero, then it quantifies a spatio-compositional inhomogeneity (SpCoIN)[1]. Notice, that when in section 2.2 we discuss the CoIN measure in the math-physical context, it should not be mistaken for another sort of chemical composition inhomogeneity. It might be interesting to mention that for one-dimensional disordered models with long-range correlations in random potentials, the structural disorder relates to randomly perturbed the spacing between the barriers while the compositional one refers to variations in strength of the barriers [17]. Generally, at a given spatial scale our measure can discriminate between four main categories of spatio-compositional inhomogeneity: o-o, o-d, d-o and d-d, where on the first (second) place the letters "o" and "d" describe spatial (compositional) order and disorder, respectively.

The present SpCoIN measure can be utilized as a complementary tool to traditional ones for analysis of multiscale variability at discrete time scales of any type of time series (even so-called "gappy" ones [18]) with integer valued genuine data or transferable to integer form. Among such exemplary systems one can find intriguing in climatology Southern Oscillation Index [19], fractional Brownian motion analysed recently by Tsallis permutation entropy [20], coarse-grained heart rate data investigated by multiscale entropy and multiscale time irreversibility methods [21], growth models with grains at the film surfaces [22,23], model of surface relaxation with interesting dynamics [24] belonging to so-called zero-range processes reviewed in Ref. [25].

For two-dimensional systems, any 8-bit greyscale image (also a colour pattern carefully converted to the greyscale) can be a subject of the multiscale analysis. Interesting opportunities produce theoretical patterns obtained for epidemic models with spatial structure based on the cellular automata method [26], ratio-dependent predator-prey model [27], parametrically forced patterns and quasipatterns [28,29] or a hybrid spin-1 Ising model with non-local constraints imposed by the Bak–Tang–Wiesenfeld sandpile model of self-organized criticality, where ''grain heights'' $h_l \in \{0,1,2\}$ are special variables [30], to mention just a few of them.

The method advanced in this paper accounts for a competition of two kinds of disorder in FSO-systems. This idea is reminiscent of the fruitful approach of competing interactions used in physics and one can expect wide-ranging applications of our method. The basic formula is given below in Eq. (8) while its version for a system additionally constrained, in Eq. (11). We would like to emphasize that the two-component entropic measure for multiscale analysis of SpCoIN is fairly flexible as will show given examples of quite different systems of FSOs.

## 2. Model system of three-dimensional pillars

In this work, we study a model system of three-dimensional pillars of size $1 \times 1 \times h_l$ composed of indistinguishable unit cubes. As for application purposes the pillar heights can be identified with shades of grey. The pillar decomposability strongly influences both the microcanonical entropy and the highest its possible value, which are needed to construct the final measure. The case when every pillar is treated as the whole entity deserves a separate research, see formula (A.1) given in Appendix. We assume that at every length scale the number of pillars and sum of their heights conserves when the entropy is maximized. The considered two-component measure deals with a spatio-compositional inhomogeneity. In addition to spatial degrees of freedom (spatial dof) also a different type associated with feasible compositions of a cell sum of pillar heights exactly into a given number of pillars at each cell is involved. The latter type is named compositional dof. Thus, besides the pure spatial inhomogeneity of pillar locations in each cell, a kind of compositional inhomogeneity of cell sums of their heights can be also quantified, separately or together.

### 2.1 Actual and reference macrostates

For the sake of clarity a window of width $k$ we identify with the one-dimensional cell of length $k$ while for patterns the cell is of size $k \times k$. In the following we shall simply use the notion "cell" that size be clear from the system's dimension $d = 1, 2$. To overcome the limitation of standard pattern partitions (into non-overlapping cells) to the scales, for which $k$ is an integer divisor of $L$ here we use a sliding cell-sampling (SCS) approach [7]. To control cell statistics a simple condition for this method is also further given in section 2.4.

In general, given time series of length $L$ or pattern of size $L \times L$ can be sampled by $\kappa = [(L - k)/z + 1]^d$ cells with a sliding factor $1 \leq z \leq k$ provided $(L - k) \mod z = 0$. Here the $z = 1$ is chosen that gives the maximal overlapping of the cells. Notice that for $z = k$ there is no overlapping and the standard partitioning is recovered. In fact, in this way we analyse auxiliary series $L_a$ or pattern $L_a \times L_a$, where $L_a \equiv [(L-k)/z + 1]\, k$. Such auxiliary representative systems composed of the sampled cells placed in a non-overlapping manner clearly reproduce the general structure of the initial ones, cf. Fig. 1(b) in Ref. [7,13]. Instead of preferred here Greek letter $\kappa$, the notation $\chi_a$ has been used in Ref. [7].

One can envisage a grey level pattern evolving in time according to given rules like the recently investigated approximate quasipatterns [28,29]. However, at this stage the systematic investigation of time evolution itself for systems is omitted. As it was mentioned above, in language of our present model the pillar heights being positive integer numbers can be matched with the shifted grey levels of the range 1−256 at every moment in time. The zero value corresponds to non-occupied positions by pillars or grey level values. Such positions can be linked with the missing or incomplete data case. Generally, during the sampling procedure $i$th non-empty cell can be occupied by $n_i(k)$ pillars of heights $h_l(i, k) \geq 1$ on positions $l = 1, 2, \dots, n_i(k) \leq k^d$. Then, for the corresponding local sum $H_i(k) \equiv \sum_{l=1}^{n_i} h_l(i, k)$ of the pillar heights we have always $H_i(k) \geq n_i(k) \geq 1$. Only for empty $i$th cell $n_i(k) = H_i(k) = 0$. Keeping this in mind the natural constraints for every length scale $k$ can be written as:

$$(i) \quad \sum_{i=1}^{\kappa} n_i(k) = n(k), \qquad (ii) \quad \sum_{i=1}^{\kappa} H_i(k) = H(k). \tag{1}$$

To simplify notation we will omit the parameter $k$ wherever it doesn't lead to misunderstanding. When in Eq. (1) condition $(ii)$ only is taken into account, the definition of a configurational macrostate at a given length scale $k$ employs a set $\{H_i(k)\}$ alone. To avoid confusion, in the present model we use the letters $H_i$ and $H$ for the local and total sum of non-zero pillar heights instead of the letters $G_i$ and $G$ for the local and total sum of grey level values 0−255 used in Ref. [13].

According to [31], with the usage of present notation, a composition of nonnegative integer $H_i$ into exactly $n_i$ positive integers $h_i$ or $n_i$-composition is any solution $(h_1, h_2, \dots, h_{n_i})$ of $h_1 + h_2 + \cdots + h_{n_i} = H_i$ with integer $h_i \geq 1$, $i \in \{n_i\}$ (the order of the summands counts). The number of $n_i$-compositions of $H_i$ is simply $\binom{H_i - 1}{n_i - 1}$. Continuing, a composition where some of the $h_i$ are allowed to be zero is called a weak $n_i$-composition of $H_i$ [32]. Exactly weak $k^2$-compositions of $G_i$ were considered for recently proposed versatile entropic measure (VEM) of grey level inhomogeneity [13]. The number $\Omega_{\text{ext}}$ of realizations for the simplest macrostate $\{G_i(k)\}$ there was calculated according to the formula

$$\Omega_{\text{ext}}(k, G) = \prod_{i=1}^{\kappa} \binom{G_i + k^2 - 1}{k^2 - 1}. \tag{2}$$

The present model besides the set $\{H_i\}$ incorporates also the set $\{n_i\}$ for each $i$th cell. This leads to significant changes in definitions of generalized macrostates and the corresponding numbers of microstates. However, we show in section 3 that the current generalized approach includes the previous measure described in Ref. [13].

To access the SpCoIN we shall apply microcanonical entropy $\tilde{S}(k) = k_B \ln \tilde{\Omega}$, where the Boltzmann constant will be set as $k_B = 1$ for convenience and $\tilde{\Omega}$ denotes the number of equally probable possible microstates for generalized actual and reference configurational macrostates. Now, actual macrostate AM($k$) can be define by the corresponding set, $\{n_i(k), H_i(k)\}_{\text{AM}}$ with $i = 1, 2, \dots, \kappa$, where the cell occupation numbers $n_i(k; \text{AM})$ and local sums $H_i(k; \text{AM})$ of pillar heights are obtained on the basis of existent pattern. Clarifying further notation, let $\omega_1(i)$ be a set of possible arrangements of $n_i$ pillars (or $n_i$-arrangements) in the base plane at $i$th cell. Such a set can be equivalently called spatial dof at $i$th cell. Let $\omega_2(i)$ stands for the set of all $n_i$-compositions of $H_i$. It can be respectively called compositional dof at $i$th cell. Now, by $f_i$ we denote a product of $\omega_1(i)$ and $\omega_2(i)$. Taking into account the constraints (*i*) and (*ii*) given by Eq. (1) one can calculate the number $\widetilde{\Omega}$ of microstates for AM

$$\widetilde{\Omega}(k,n,H) = \prod_{i=1}^{\kappa} \omega_1(i)\,\omega_2(i) \equiv \prod_{i=1}^{\kappa} f_i\,, \tag{3}$$

where

$$f_i = \begin{cases} \binom{k^d}{n_i} \equiv \omega_1(i) & \text{for cell occupied by trivial pillars of unit height } -\text{(a)} \\ \binom{k^d}{n_i}\binom{H_i-1}{n_i-1} \equiv \omega_1(i)\omega_2(i) & \text{for partially occupied cell by nontrivial pillars } -\text{(b)} \\ \binom{H_i-1}{k^d-1} \equiv \omega_2(i) & \text{for fully occupied cell by nontrivial pillars } -\text{(c)} \\ 1 & \text{for empty cell}\,. \end{cases} \tag{4}$$

The attributed letters (a-c) refer to the classes of model macrostates. A toy model discussed in sections 2.2 and 2.3 exemplifies those macrostates in context of possible forms of the entropic measure and its maximization. The case (a) we shall neglect here as equivalent one to binary patterns considered previously [6,7]. For the cases (b) and (c), from Eqs. (4) and (5) one can clearly specify the numbers $\widetilde{\Omega}(k,n,H;\text{b}) \equiv \widetilde{\Omega}(\text{b})$ and $\widetilde{\Omega}(k,n,H;\text{c}) \equiv \widetilde{\Omega}(\text{c})$ for the corresponding AM($k$; b) and AM($k$; c).

On the other hand, to obtain the necessary for our measure the highest possible value of the entropy, $\widetilde{S}_{\max}(k) = \ln \widetilde{\Omega}_{\max}$, we need a reference macrostate RM($k$). It is defined by the appropriate set, $\{n_i(k), H_i(k)\}_{\text{RM}}$ with $i = 1, 2, \ldots, \kappa$. Now, the integer numbers $n_i(k; \text{RM})$ and $H_i(k; \text{RM})$ at a given length scale $k$ and for fixed $n(k)$ and $H(k)$ should ensure the largest number $\widetilde{\Omega}_{\max}$ of realizations for this RM. However, to the best knowledge of the author, no explicit formula valid *generally* can be provided for RM in our pillar model. For RM(a) and RM(c) it can be obtained through pure theoretical analysis with usage of Lagrange multiplier method. The highest possible value of the entropy $\widetilde{S}_{\max}(k; \text{c})$ relates to a set of configurations of local sums of pillar heights distinguished by a sufficient (for this case) simple condition $\left| H_i(k; \text{RM}) - H_j(k; \text{RM}) \right| \le 1$, holding for each pair of fully occupied cells. For case (a) the similar form of the condition but with cell occupation numbers $n_i$ has been already applied to binary patterns [6,7]. In the present case (c) the desired formula for the number $\widetilde{\Omega}_{\max}(\text{c})$ of microstates for RM(c) can be written as

$$\widetilde{\Omega}_{\max}(k,n,H;\text{c}) = \binom{H_0-1}{k^d-1}^{\kappa-R_0} \binom{H_0}{k^d-1}^{R_0}, \tag{5}$$

where $R_0 = H \bmod \kappa$ and $H_0 = (H - R_0)/\kappa$. Every microstate of the set $\widetilde{\Omega}_{\max}(k,n,H;\text{c}) \equiv \widetilde{\Omega}_{\max}(\text{c})$ represents a reference macrostate $\{n_i = k^d, H_i \in [H_0, H_0+1]\}_{\text{RM}}$ with $\kappa - R_0$ and $R_0$ fractions of cells with local sums $H_0$ and $H_0 + 1$, respectively.

For RM(b), on the basis of rational arguments and computer simulations, its numerical finding is still available. At given length scale $k$, under assumed integer ranges for $i \in [-\alpha, \ldots, \beta]$ and $j \in [-\gamma, \ldots, \delta]$ the appropriate solutions can be find for the sets of integer coefficients $\{x\} \equiv (x_{-\alpha}, \ldots, x_0, \ldots, x_\beta)$ and $\{y\} \equiv (y_{-\gamma}, \ldots, y_0, \ldots, y_\delta)$ fulfilling the basic equations:

$$x_{-\alpha}(n_0-\alpha) + \ldots + x_{-1}(n_0-1) + x_0 n_0 + x_1(n_0+1) + \ldots + x_\beta(n_0+\beta) = n\,, \tag{6a}$$

$$y_{-\gamma}(H_0-\gamma) + \ldots + y_{-1}(H_0-1) + y_0 H_0 + y_1(H_0+1) + \ldots + y_\delta(H_0+\delta) = H, \tag{6b}$$

and

$$x_{-\alpha} + \ldots + x_{-1} + x_0 + x_1 + \ldots + x_\beta = \kappa, \tag{7a}$$

$$y_{-\gamma} + \ldots + y_{-1} + y_0 + y_1 + \ldots + y_\delta = \kappa, \tag{7b}$$

where $r_0 = n \bmod \kappa$ and $n_0 = (n - r_0) / \kappa$; while $R_0$ and $H_0$ are already defined by Eq. (5). Then, among obtained candidates for RM(b) one can select that one (or those ones in the case of a degeneration) with the highest possible value of the entropy $\tilde{S}_{max}(k; b)$. However, the optimal ranges for $i, j$ indices depend on a system's size and its parameters.

For instance, in sections 2.2 and 2.3 we briefly discuss a toy model that can be used to show what the RM(b) and RM(c) are. For the number of pillars $n = 9$ the sets $\{x\} \equiv (x_0, x_1)$ and $\{y\} \equiv (y_{-4}, \ldots, y_0, \ldots, y_2)$ are needed while for $n = 19$ the sets $\{x\} \equiv (x_0, x_1)$ and $\{y\} \equiv (y_{-1}, y_0, y_1)$, see Tab. 1 in section 2.2. On the other hand, the shadowed bottom row in Tab. 1 for $n = 20$ exemplifies the properties of (5). We can see that the corresponding RM(c) relates to the sets $\{x\} \equiv (x_0 = 5)$ and $\{y\} \equiv (y_0 = \kappa - R_0 = 3, y_1 = R_0 = 2)$, which are connected with Eqs. (6a, b) and (7a, b) for $n_0 = 4$, $r_0 = 0$, $H_0 = 5$ and $R_0 = 2$. To be more flexible the sketched above general approach requires further improvements, which are underway.

## 2.2 Two-component entropic measure

To characterize the *relative* spatio-compositional inhomogeneity we have to quantify the statistical dissimilarity of AM and RM macrostate at every length scale $1 \le k \le L$. For its reasonable comparison at different scales a difference of the corresponding entropies per cell is used. Therefore, a two-component multiscale entropic measure of SpCoIN can be written as

$$
\begin{aligned}
\tilde{S}_\Delta(k, n, H) &= [\tilde{S}_{max}(\text{RM}) - \tilde{S}(\text{AM})] / \kappa \\
&= [(S_{(max)} + G_{(max)}) - (S + G)] / \kappa \\
&= [(S_{(max)} - S) + (G_{(max)} - G)] / \kappa \equiv S_\Delta(\text{I}) + G_\Delta(\text{II}).
\end{aligned}
\tag{8}
$$

The first component, $S_\Delta(\text{I})$, contains difference of $S_{(max)}$ and $S = \Sigma \ln \omega_1(i)$ per cell, and corresponds to spatial inhomogeneity. The second one, $G_\Delta(\text{II})$, includes difference of $G_{(max)}$ and $G = \Sigma \ln \omega_2(i)$ per cell, and refers to compositional inhomogeneity. Note, that the subscripts in brackets indicate that both terms $S_{(max)}$ and $G_{(max)}$ describe in general maximal value of $\tilde{S}_{max} = S_{(max)} + G_{(max)}$, i.e. maximum of their sum, not necessary maximums each of them separately. According to definition at boundary length scales we have $\tilde{S}_\Delta(k = 1) = \tilde{S}_\Delta(k = L) = 0$. Symbolically, one can write SpCoIN = SpIN + CoIN.

The maximums (minimums) of the measure indicate those scales at which higher (lower) average SpCoIN appears compared to neighbour scales. Let us focus on the first maximum (usually clearly higher than the others), when it is assumed that SpIN = 0. Then, a peak of the CoIN quantified by $G_\Delta(\text{II})$ can be reasonable interpreted as an indicator of formation of specific 'height clusters' of pillars. Remembering the natural constraints given by Eq. (1), one can expect that at this scale denoted as $k_{max}$, the characteristic clustering in heights *increases* the number of cells both with large and small values of local sums $H_i(k_{max})$ of pillar heights in comparison with those expected for a realizable maximally uniform distribution of pillar with average height close to $H_0(k_{max}) \approx H_0(k_{max}) + 1$. On the other hand, the equally distant and comparable in values minimums of the entire measure $\tilde{S}_\Delta(k, n, H)$ relate to the *statistical* spatio-compositional periodicity.

Consider now the behaviour of the two-component entropic SpCoIN measure in connection with the possible categories of generalized macrostates. We can distinguish the following three their categories:

a) $n = H \le \kappa k^d$.

All pillars become trivial since they are of unit height. At every $i$th cell the compositional dof reduce to $\omega_2(i) = 1$ and the $f_i$-product defined early by Eqs. (3) and (4) simplifies to $\omega_1(i)$. Correspondingly, from (8) results $\tilde{S}_\Delta(\text{a}) \equiv S_\Delta(\text{I}) \rightarrow S_\Delta = [S_{max} - S] / \kappa$. The pure measure $S_\Delta$ corresponds to the binary one already investigated [6,7], where spatial inhomogeneity (now abbreviated as SpIN) for patterns of black and white pixels can be linked just to the sets $\{n_i\}$ and $\{n_i\}_{\text{RM}}$.

b)  $n < \kappa k^d$  and  $n < H$.

This intermediate case for incomplete, missing or gappy data is the most complex one. Now, at least some of cells are partially occupied by non-trivial pillars. According to Eqs. (3) and (4) for those cells both $\omega_1(i)$ and $\omega_2(i)$ contribute to the two-component entropic measure (8), $\tilde{S}_{\Delta}(\text{b}) = S_{\Delta}(\text{I}) + G_{\Delta}(\text{II})$, for the SpCoIN that in this case is connected with the full sets $\{n_i, H_i\}$ and $\{n_i, H_i\}_{\text{RM}}$.

c)  $n = \kappa k^d < H$.

There are no empty positions for pillars in the base plane. At every $i$th cell the spatial dof reduce to $\omega_1(i) = 1$ and the $f_i$-product given by Eqs. (3) and (4) equals to $\omega_2(i)$. Consequently, from (8) we obtain $\tilde{S}_{\Delta}(\text{c}) \equiv G_{\Delta}(\text{II}) \to G_{\Delta} = [G_{\max} - G] / \kappa$. The pure measure $G_{\Delta}$ refers to CoIN that connects with the sets $\{H_i\}$ and $\{H_i\}_{\text{RM}}$. Its simplified counterpart (VEM) of grey level inhomogeneity was introduced in Ref. [13].

A pictorial diagram in Fig. 1 illustrates above categories on example of a one-dimensional toy model discussed also in the next section. This model contains $\kappa = 5$ cells of size $k = 4$ while a given total sum of pillar heights $H \geq 1$ and total number of pillars $n \leq 20$. Each kind of the symbols in the diagram indicates a one of three allowed categories of generalized macrostates for $H = 1, \ldots, 27$ on dependence of $n = 1, \ldots, 20$.

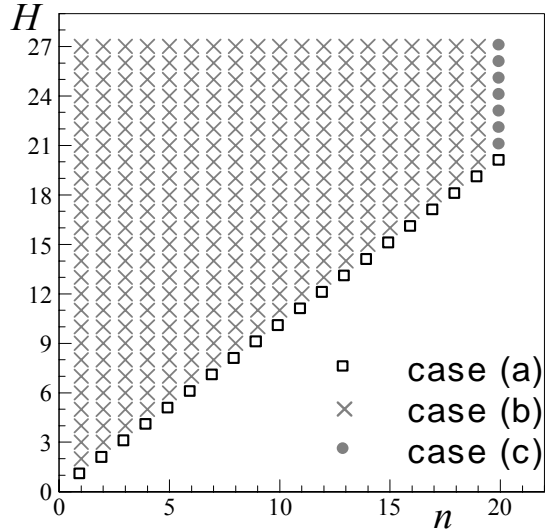

**Fig. 1.** Diagram of allowable categories of the macrostates for a one-dimensional toy model with a fixed number $\kappa = 5$ of cells of size $k = 4$ and changing total sum of pillar heights $1 \leq H \leq 27$ on dependence of total number of pillars $1 \leq n \leq 20$. (a) $n = H \leq \kappa k^d$. (b) $n < \kappa k^d$ and $n < H$. (c) $n = \kappa k^d < H$.

### 2.3. Maximization of entropy of a toy model

Our toy model can be also used to show what are the maximal two-component entropies $\tilde{S}_{\max}(k = 4, n, H = 27)$ as a function of total number $n$ of pillars for given RM(b) and RM(c). The RMs were determined by a computer checking of all possible configurations. The obtained RMs we collect in Tab. 1.

**Table 1**. The reference macrostates $\{n_i, H_i\}_{RM}$, $i = 1,\dots,5$, and the entropic measure $\tilde{S}_{max}(n)$ for a one-dimensional toy model with a fixed number of cells $\kappa = 5$ of size $k = 4$ and constant total sum of pillar heights $H = 27$ (cf top row on pictorial diagram in Fig. 1) for different total number of pillars $n = 1,\dots,20$.

| $n$ | $N_1,H_1$ | | $n_2,H_2$ | | $n_3,H_3$ | | $n_4,H_4$ | | $n_5,H_5$ | | $\tilde{S}_{max}$ |
|---|---|---|---|---|---|---|---|---|---|---|---|
| 1 | 0 | 0 | 0 | 0 | 0 | 0 | 0 | 0 | 1 | 27 | 0.2773 |
| 2 | 0 | 0 | 0 | 0 | 0 | 0 | 0 | 0 | 2 | 27 | 1.0100 |
| 3 | 0 | 0 | 0 | 0 | 0 | 0 | 0 | 0 | 3 | 27 | 1.4340 |
| 4 | 0 | 0 | 0 | 0 | 0 | 0 | 2 | 13 | 2 | 14 | 1.7267 |
| 5 | 0 | 0 | 0 | 0 | 0 | 0 | 2 | 9 | 3 | 18 | 2.0340 |
| 6 | 0 | 0 | 0 | 0 | 2 | 9 | 2 | 9 | 2 | 9 | 2.3227 |
| 7 | 0 | 0 | 1 | 1 | 2 | 8 | 2 | 9 | 2 | 9 | 2.5733 |
| 8 | 0 | 0 | 2 | 6 | 2 | 7 | 2 | 7 | 2 | 7 | 2.8304 |
| 9 | 1 | 1 | 2 | 6 | 2 | 6 | 2 | 7 | 2 | 7 | 3.0711 |
| 10 | 2 | 5 | 2 | 5 | 2 | 5 | 2 | 6 | 2 | 6 | 3.2673 |
| 11 | 2 | 4 | 2 | 4 | 2 | 5 | 2 | 5 | 3 | 9 | 3.3711 |
| 12 | 2 | 4 | 2 | 4 | 2 | 4 | 3 | 7 | 3 | 8 | 3.4393 |
| 13 | 2 | 4 | 2 | 4 | 3 | 6 | 3 | 6 | 3 | 7 | 3.4506 |
|  | 2 | 3 | 2 | 4 | 3 | 6 | 3 | 7 | 3 | 7 | 3.4506 |
|  | 2 | 3 | 2 | 3 | 3 | 7 | 3 | 7 | 3 | 7 | 3.4506 |
| 14 | 2 | 3 | 3 | 6 | 3 | 6 | 3 | 6 | 3 | 6 | 3.4481 |
| 15 | 3 | 5 | 3 | 5 | 3 | 5 | 3 | 6 | 3 | 6 | 3.3824 |
| 16 | 3 | 5 | 3 | 5 | 3 | 5 | 3 | 5 | 4 | 7 | 3.1416 |
| 17 | 3 | 5 | 3 | 5 | 3 | 5 | 4 | 6 | 4 | 6 | 2.8279 |
|  | 3 | 4 | 3 | 5 | 3 | 5 | 4 | 6 | 4 | 7 | 2.8279 |
|  | 3 | 4 | 3 | 4 | 3 | 5 | 4 | 7 | 4 | 7 | 2.8279 |
| 18 | 3 | 4 | 3 | 4 | 4 | 6 | 4 | 6 | 4 | 6 | 2.5141 |
|  | 3 | 4 | 3 | 5 | 4 | 6 | 4 | 6 | 4 | 6 | 2.5141 |
| 19 | 3 | 4 | 4 | 6 | 4 | 6 | 4 | 6 | 4 | 6 | 2.1558 |
| 20 | 4 | 5 | 4 | 5 | 4 | 5 | 4 | 6 | 4 | 6 | 1.7528 |

For a few cases a kind of degeneration of the RMs is revealed. For instance, $\tilde{S}_{max}(n = 13)$ reaches its highest value equal to 3.4506 within the range of allowed $n$ for $g = 3$ different RMs. In turn, for $n = 17$ and 18 the corresponding degeneration factors are equal to $g = 3$ and 2, cf Tab. 1. We expect that for much larger systems such kind of degeneration should become a very rare event.

The corresponding maximal two-component entropy $\tilde{S}_{max}$ and its parts $S_{(max)}$ and $G_{(max)}$ are shown in Fig. 2 as functions of $n$. For various $n$ one can observe how the spatial dof compete with its compositional counterpart for maximizing the $\tilde{S}_{max}(4, n, 27)$. It should be stressed that for $n < 20$ we have the most general case (b). Despite of simplicity of our toy model, for certain numbers of pillars we have $S_{(max)} < G_{(max)}$ while for the others the opposite inequality holds, $S_{(max)} > G_{(max)}$. This means that even for a fixed total sum of pillar heights $H$ the spatial disorder and the compositional one can affects the entropy $\tilde{S}_{max}$ in a different degree.

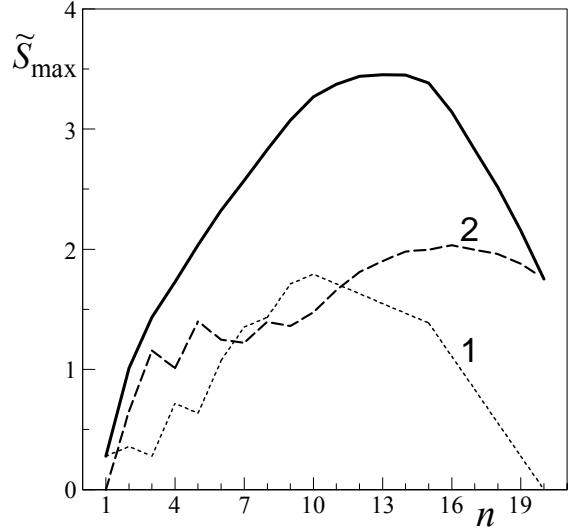

**Fig. 2.** Maximal two-component entropy $\tilde{S}_{max} = S_{(max)} + G_{(max)}$ and the competing for its maximizing the two terms, $1$ – the spatial $S_{(max)}$ and $2$ – the compositional $G_{(max)}$, as functions of the total number $n$ of pillars for the 1D toy model discussed in the text and under the same conditions as described in Tab. 1. Here, the peak of $\tilde{S}_{max}$ appears for $n = 13$ (cf also the corresponding degenerated RMs given in Tab. 1).

### 2.4  $h^*$-constrained two-component entropic measure

The generalized entropic measure given by Eq. (8) is the exact one if the only upper limit for the maximal height of pillars comes from the two natural conservation rules (*i*) and (*ii*) described by Eq. (1)

$$h_{\max}(b,c) = \begin{cases} H - n + 1 & \text{for case (b)} \\ H - \kappa k^d + 1 & \text{for case (c)} \end{cases} \tag{9}$$

However, as we see from a one of examples presented in the next section (cf Fig. 3) the $\tilde{S}_\Delta$ becomes the qualitatively correct measure for systems with a more demanding additional $h^*$-constraint for the height $h$ of pillars

$$(iii) \qquad h \le h^* < h_{\max} . \tag{10}$$

For instance, for grey level values shifted to the range $1-256$ as it was mentioned early, the additional constraint equals to $h^* = 256$. When $h^* < h_{\max}(b, c)$, for the sets $\tilde{\Omega}(b)$ and $\tilde{\Omega}(c)$ compared with the $\tilde{\Omega}(b; h^*)$ and $\tilde{\Omega}(c; h^*)$ the former ones include some excess compositional microstates at least, i.e. $\tilde{\Omega}(b; h^*) \in \tilde{\Omega}(b)$ and $\tilde{\Omega}(c; h^*) \in \tilde{\Omega}(c)$. Therefore, only the $h^*$-constrained entropic measure is the *exact* one in such a case. Its general form can be written as

$$\begin{aligned} \tilde{S}_\Delta(k,n,H;h^*) &= \left[ \tilde{S}_{\max}(\text{RM};h^*) - \tilde{S}(\text{AM};h^*) \right] / \kappa \\ &\equiv S_\Delta(\text{I};h^*) + G_\Delta(\text{II};h^*) , \end{aligned} \tag{11}$$

where the two $h^*$-constrained components $S_\Delta(\text{I}; h^*)$ and $G_\Delta(\text{II}; h^*)$ are defined as in Eq. (8). To distinguish the $h^*$-constrained entropic measure given by Eq. (11) from the $h^*$-unconstrained its counterpart described by Eq. (8) we will use the abridged notation for them, $\tilde{S}_\Delta(h^*)$ and $\tilde{S}_\Delta$, respectively.

To the best knowledge of the author, in general the number of microstates for $h^*$-constrained macrostates does not possess a closed-form expression. To obtain the proper numbers, $\tilde{\Omega}_{\max}(c, k; h^*)$ and $\tilde{\Omega}(c, k; h^*)$, of the microstates we are forced to use the combinatorial recipe [33]. It states, that the number of *restricted* compositions of $N$ indistinguishable elements (here unit cubes) with exactly $r$ summands (here $n_i$ pillars that order counts) under condition that the number $x$ of its elements (here the height $h_i$ of each pillar) belongs to the interval $q \le x \le q + s - 1$ (here $q \equiv 1$ and $s \equiv h^*$) equals to coefficient $a_p$ of $x^p$ with $p \equiv N - qr$ for an expansion of the generating function

$$\left( \frac{1-x^s}{1-x} \right)^r = (1 + x + x^2 + \dots + x^{s-1})^r \equiv \sum_{p=0}^{r(s-1)} a_p x^p . \tag{12}$$

For instance, at a given length scale $k$ for case (c) the number $N$ takes value $H_i$ at $i$th cell for $\tilde{\Omega}(c, k; h^*)$, $H_0$ and $H_0 + 1$ for the appropriate fractions of cells $\kappa - R_0$ and $R_0$ for $\tilde{\Omega}_{\max}(c, k; h^*)$. As an illustrative example let us consider the $h^*$-constrained number of compositions of $H_i = 7$ into exactly $n_i = 3$ summands for $h^* = 4$. Following the recipe given by Eq. (12) we have $a_4 = 12$ while the total number of compositions equals to 15. It can be additionally divided into four classes: $\{(5,1,1), (1,5,1), (1,1,5)\}$ = I-class, $\{(4,2,1), (4,1,2), (2,4,1), (1,4,2), (2,1,4), (1,2,4)\}$ = II-class, $\{(3,3,1), (3,1,3), (1,3,3)\}$ = III-class and $\{(3,2,2), (2,3,2), (2,2,3)\}$ = IV-class. Indeed, only twelve configurations among them belonging to the last three classes fulfil the above constraint.

It should be stressed that the recipe described by Eq. (12) is not easily implemented. For $h^*$ typical for 2D grey level patterns the recurrence procedure for the coefficients $a_p$ needs large memory and is also time-consuming. However, we demonstrate numerically in section 3 (cf Fig. 3) for the most common case (c) that the $h^*$-constraint is not destructive yet for a nice correlation of the $\tilde{S}_\Delta(c; h^*)$ and $\tilde{S}_\Delta(c)$, at least for 1D systems. In general, such a correlation can be also roughly understood. Let us suppose for the corresponding macrostates that their behaviour is close to the proportional one, i.e., $\hat{\Omega}(c, k; h^*) \approx \gamma(k)\hat{\Omega}(c, k)$ and $\hat{\Omega}_{max}(c, k; h^*) \approx \gamma_{max}(k)\hat{\Omega}_{max}(c, k)$, where the coefficients $0 < \gamma(k)$, $\gamma_{max}(k) < 1$ depend only on the length scale $k$. Thus, the $\tilde{S}_\Delta(c, k; h^*) - \tilde{S}_\Delta(c, k) = \ln[\gamma_{max}(k)/\gamma(k)]/\kappa$ is a common function of $k$ at different length scales and the appearance of such a correlation is not surprising. For the systems compared in Fig. 3 in section 3, the difference of the entropic measures is positive what implies $\gamma_{max}(k) > \gamma(k)$ in that case.

Therefore, when condition (*iii*) given by Eq. (10) applies, instead of the $\tilde{S}_\Delta(c, k; h^*)$ the usage of its qualitatively correct and the less demanding counterpart $\tilde{S}_\Delta(c, k)$ is much easier, cf Figs. 4 and 5 in section 3. Nevertheless, at least for the initial length scales $k < 32$ even for 2D greyscale images the most general two-component $h^*$-constrained entropic measure can be calculated on a standard personal computer (PC) provided we employ (earlier prepared under, e.g., *Maple*) the corresponding list of all possible values of $\ln\hat{\Omega}_{max}(c, k; h^*)$ and $\ln\hat{\Omega}(c, k; h^*)$.

Last remark refers to the SCS method that involves a certain averaging process since some pillars are common for the neighbouring positions of a sliding cell. This provides rather smooth but still useful the SpCoIN characteristics over entire range of the length scales. However, for the scales $k$ close to $L$ the number $\kappa$ of sampled cells may be not large enough in comparison with the number $\kappa^*$ suitable for a good cell statistics. Therefore, for a given number $\kappa^* \leq \kappa$ one can always use the simple criterion

$$k \leq \begin{cases} L - \kappa^* + 1 & \text{for } d = 1 \\ L - \sqrt{\kappa^*} + 1 & \text{for } d = 2 \end{cases} \qquad (13)$$

to estimate the limit length scale $k$ for a given $L$, or the minimal size of a system for assumed proportion $k/L$, see examples given in Ref. [7].

## 3. Illustrative examples

For purposes of illustration, we focus on the CoIN for chosen systems. As we have matched pillar integer heights with grey levels, in this section the language of greyscale patterns is preferred. First, we point out that the present approach confirms the striking effect of intersecting $S_{ext,\Delta}$ curves observed for pairs of differently contrasted grey level patterns [13] (c.f. Figs. 1 and 2). The hidden statistical grey level periodicity detection by the equally distant minimums of the $S_{ext,\Delta}$ measure found in Ref. [7] (cf Figs. 2 and 3), can be also exactly reconstructed by the present $\tilde{S}_\Delta(c) \rightarrow G_\Delta$ measure of CoIN.

This becomes clear if we realize that in the case (c) discussed in section 2.2 we have $n_i = k^d$. Thus, when all grey level values 0–255 at each *i*th-cell are shifted up by 1 within the present approach, then every local sum $G_i$ becomes equal to $G_i + k^d = H_i$. Thus, expressing Eq. (2) on dependence of $H_i$ we get for case (c) the appropriate term $\omega_2(i)$ in Eq. (4). Therefore, when the spatial degrees of freedom are excluded, the present $G_\Delta$ measure and the previous $S_{ext,\Delta}$ are equivalent mathematically to each other. So, one can state that compositional inhomogeneity alone in language of the present model means the same as grey level inhomogeneity in the previous approach [13].

One more remark is in order. Among three examples considered below, the two time series presented in the insets of Fig. 3 are of synthetic type although linked to well-known in physics fractional Brownian motion (fBm) while in Figs. 4 and 5 the insets relate to the source files kindly provided by their authors. Those original files refer to forced evolution in time of a numerical solution of the model partial differential equation (PDE) [28,29] (devised to obtain examples of superlattice patterns and quasipatterns) and Monte Carlo reconstruction process of an initial real laser-speckle pattern [2] (the II part). In contrast to the previous work [13], the subject of the CoIN multiscale analysis are systems, say $X$ and $Y$–patterns, with considerably different total sums of non-zero grey levels, e.g., $H(X) > H(Y)$. Only the $G_\Delta(X; k)$ per grey level $\equiv G_\Delta(X; k)/H(X)$ and similarly, $G_\Delta(Y; k)/H(Y)$, guarantee a responsible comparison at a given $k$. To avoid extremely low values we simply rescale the compared measures multiplying each of them by the highest of the total sums, i.e., by the $H(X)$ in this artificial example.

In Fig. 3 the CoIN quantified by the $h^*$-constrained entropic measure $G_\Delta(h^*)$ with $h^* = 256$, dashed lines, and its counterpart the $G_\Delta$, solid lines, are compared for the related profiles each of length $L = 256$ depicted for better resolution in two parts in the insets. Both profiles were obtained by random midpoint displacement method with the same seed for a pseudo-random numbers generator but with slightly different Hurst exponents, $H_1 = 0.4$ for grey lines, and $H_2 = 0.5$ for black ones. The point is, whether the entropic measure can discern that some underlying small changes in exponents have taken place.

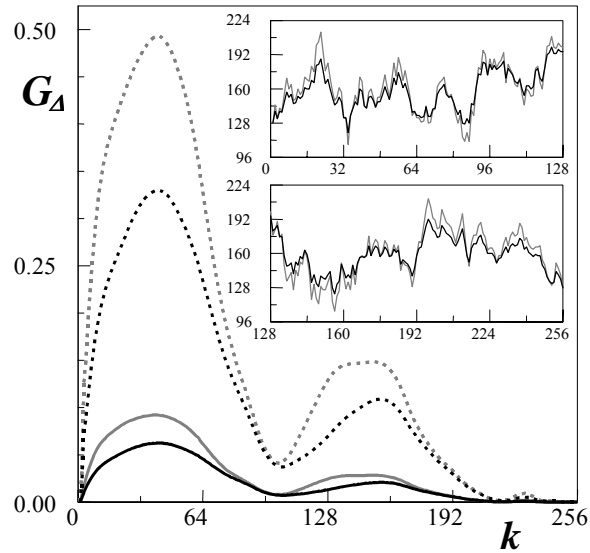

**Fig. 3.** The multiscale compositional inhomogeneity (CoIN) quantified by the entropic measure $G_\Delta(k)$, solid lines, and $G_\Delta(k; h^*)$, dashed lines, for two profiles each of length $L = 256$ for fBm, which for better resolution are shown in two parts in the insets. Those profiles with slightly different Hurst exponents equal to $H_1 = 0.4$, grey lines, and $H_2 = 0.5$, black ones, were obtained by random midpoint displacement method with the same seed. For the same Hurst exponent the both measures well correlate.

In Fig. 3 both variants of the measure, the $G_\Delta$ and $G_\Delta(h^*)$, make such a clear distinction in accordance with our expectations, i.e., $G_\Delta(H_1) > G_\Delta(H_2)$, grey and black solid lines, and $G_\Delta(h^*; H_1) > G_\Delta(h^*; H_2)$, grey and black dashed lines. The $G_\Delta(h^*)$ measure appears to be relatively the more sensitive. The position of the first peak is the same for both curves, $k_{max}(H_1) = k_{max}(H_2) = 41$. Moreover, for the pairs of appropriate curves with the same Hurst exponent, i.e., $[G_\Delta(H_1), G_\Delta(h^*; H_1)]$, solid and dashed grey lines, and $[G_\Delta(H_2), G_\Delta(h^*; H_2)]$, solid and dashed black lines, there is statistically significant linear correlation. The corresponding coefficients are $r(H_1) = 0.999681$ and $r(H_2) = 0.999699$. This means, that in the case (c) the less demanding measure $G_\Delta$ provides a suitable qualitative evaluation of the CoIN. We point out that in this example, according to condition given by Eq. (13) for assumed fairly small the limit number $\kappa^* = 100$ of sampled cells, the range of length scales $1 \leq k < 155$ guarantees the better cell statistics.

With usage of the $G_\Delta$ measure that is much simpler in applications, we focus now on 2D systems with an additional constraint given by $h^* = 256$. Avoiding mathematical details, which are not a subject of the present work, we would like to analyse only the 12-fold initial, cf Fig. 18(a), intermediate (see the accompanying animation) and 14-fold final approximate quasipatterns, cf Fig. 18(b) in Ref. [28]. The corresponding extended transient of 70,000 periods for amplitudes of Fourier modes as a function of time can be clearly seen in Fig. 19 of [28]. To obtain the quasipatterns the authors solved numerically in a square domain with periodic boundary conditions the time-dependent model PDE involving the pattern-forming field $U(x, y, t)$ being a complex-valued function and real-valued $2\pi$-periodic forcing function $f(t)$, cf Eq. (3.1) in Ref. [28] or Eq. (1) in Ref. [29]. The greyscale represents the real part of $U(x, y, t)$. The exemplary $137 \times 137$ sub-domains for 12-fold $A$–initial, without a clear symmetry $B$–transient and 14-fold $C$–final cases are depicted in the insets of Fig. 4(b). Also, the characteristic greyscale histograms related to investigated here quasipatterns of size $L \times L = 500 \times 500$ (in pixels) that is a bit smaller than the original one $512 \times 512$ are depicted.

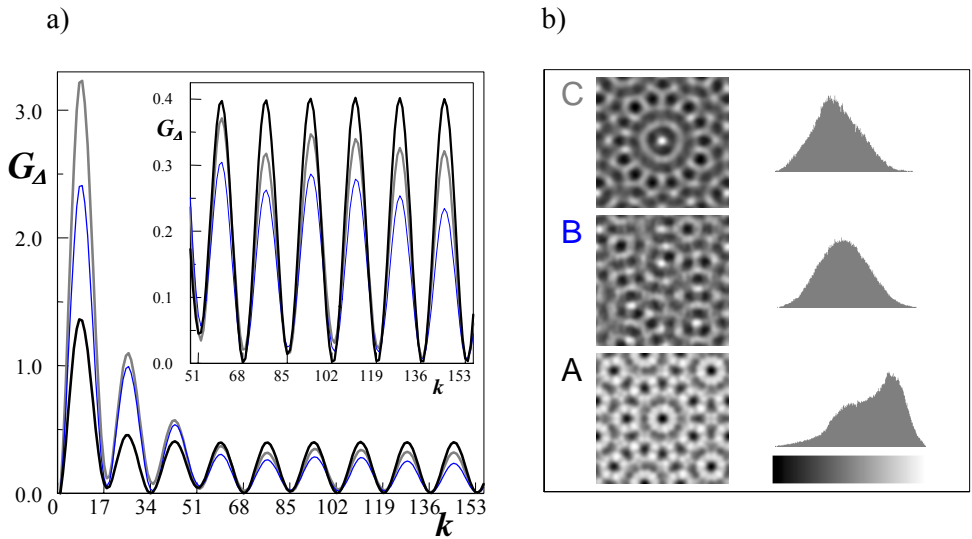

**Fig. 4.** (a) The multiscale CoIN quantified by the entropic measure $G_\Delta(k)$ per grey level for the $A$–initial (black bold line), $B$–intermediate (thin line) and $C$–final (grey line) greyscale quasipatterns of size $L \times L = 500 \times 500$ (in pixels) adapted from the animated movie of the transition from 12-fold to 14-fold case [28,29] (with the permission of the authors). One can observe around the first three peaks that $G_\Delta(A) < G_\Delta(B) < G_\Delta(C)$ while for the others peaks the weak domination of $G_\Delta(A)$ appears. Remarkably enough, the approximate *compositional* greyscale periodicity indicated by quite regular intervals between the minimums is similar for all cases. In turn, the values of the compositional inhomogeneity given by the peaks are nearly constant only for the 12-fold symmetry, see the inset; (b) The exemplary $137 \times 137$ parts of the frames from the initial (bottom) to final one (top). The characteristic greyscale histograms corresponding to the entire quasipatterns are also displayed.

For all curves, the location of the first peak (minimum) is nearly the same (the same), $k_{max} \cong 8 \div 9$ ($k_{min} = 18$), respectively. Around of the first three peaks dominates the $G_\Delta(C)$ that indicates for the highest value of the CoIN at these initial scales while the lowest value shows the $G_\Delta(A)$. This seems to be in accord with the transient process observable at the most characteristic initial length scales. Remarkably enough, for other peaks at larger scales the partially reversed behaviour can be seen with the weak domination of the $G_\Delta(A)$, see the inset of Fig. 4(a). Taking into account the hidden approximate *compositional* greyscale periodicity one can observe the following intervals between the successive local minimums, i.e., {16,17,17,17,17,17,18,17} for 12-fold $A$−initial quasipattern and {17,17,16,17,18,17,17,17} identical for both $B$−transient and 14-fold $C$−final ones within the range of the length scales $k \in (1,…,156)$.

In turn, except for the first peak position the other ones are identical for the three patterns. Let us compare also the values of the compositional inhomogeneity given by those peaks, see the inset of Fig. 4(a). They are nearly constant for case ($A$) and non-monotonously changing but correlated for cases ($B$) and ($C$). Summarizing, from the viewpoint of CoIN the above observations point to a greater similarity between the patterns ($B$, $C$) compared to the pairs ($A$, $B$) and obviously ($A$, $C$). However, as it was remarked in Ref. [28] (cf Tab. 2) the most accurate approximations to true quasipatterns depend on the exact choice of computational domain size. Thus, at this stage our observations based on the particular set of approximate quasipatterns may be rather slightly modified by the future examples of more accurate quasipatterns of these types. On the other hand, we found for a 12-fold quasipattern obtained as numerical solution of a simple rotationally-invariant model equation with only two forcing frequencies, cf Eq. (4) and Fig. 2(d) in Ref. [34], the similar general and specific features like, e.g., the decreasing (but slightly higher than others) the values of the first, third and fifth minimums of the $G_\Delta$ measure.

Before the presentation of the last example we would like to mention that the CoIN measure the $G_\Delta$ can be also used to two-valued images encoded in the shifted fashion, i.e. black pixel with grey level index $i = 1$ and white pixel with $j = 256$. In this way the sensitivity of the $G_\Delta$ can be enhanced when compared to the full greyscale counterpart. Somewhat similar "binarization" procedure has been already used to the entropic descriptor $C_\Delta(k)$ of a complex behaviour [14]. The current version of this procedure we apply to the two particularly interesting binary patterns adapted from Ref. [2] (the part II) with the permission of the authors. The initial one is a binary laser-speckle pattern shown in the left inset of Fig. 5. This pattern characterizes three-scales at least structure with compact clusters and stripes of different shapes and single black pixels described as three structural elements: "particles", "stripes" and a background "noise" in Ref. [2]. The second one is the reconstructed [2] speckle pattern shown in the right inset of Fig. 5. The authors used a specialised algorithm that employs two-point correlation functions $S_2(r)$ of the initial and target medium. Finally, all the structural elements in the reconstructed pattern are mixed in such a way that instead of three characteristic length scales a single-scale structure is generated [2]. This suggests the $S_2$-algorithm that uses limited configurational information cannot reproduce the multiscale pattern accurately in spite of the both $S_2$-functions, after finishing of the reconstruction process, are nearly identical to a very high accuracy.

The above observation [2] is quantitatively confirmed by the discrepancies in the values of the $G_\Delta(k)$ curves still present between the initial pattern (upper black bold line) and the reconstructed one (upper grey line), cf Fig. 5. Additionally, to make a comparison between the entropic measures of compositional and spatial inhomogeneity one can use a kind of binary encoding (1-black, 0-white) for the initial and reconstructed patterns. Thus, we analyse also the multiscale SpIN, compare the bottom black and grey lines for the corresponding $S_\Delta(k)$ in Fig. 5. Although they also reveal certain discrepancies, which are similar to the previous ones at larger scales, a one point is worth noticing. Namely, at the small length scales, which are structurally the most susceptible, for the initial complex pattern the $G_\Delta(k)$ measure (also the $C_\Delta(k)$) detects a clear peak at length scale $k = 6$ (7), invisible for the $S_\Delta(k)$. This peak supports the suggestion that the initial pattern is more complex, in particular at small length scales, than the reconstructed one. This underlines the usefulness of the relatively more sensitive the $G_\Delta$ measure. On the other hand, the $G_\Delta$ and $S_\Delta$ measures although of different origins, are still correlated at larger scales, say for $k > 40$ in this case. The similar behaviour shows the corresponding compositional and spatial variants of the $C_\lambda(k)$ measure, cf thin lines in Fig. 5.

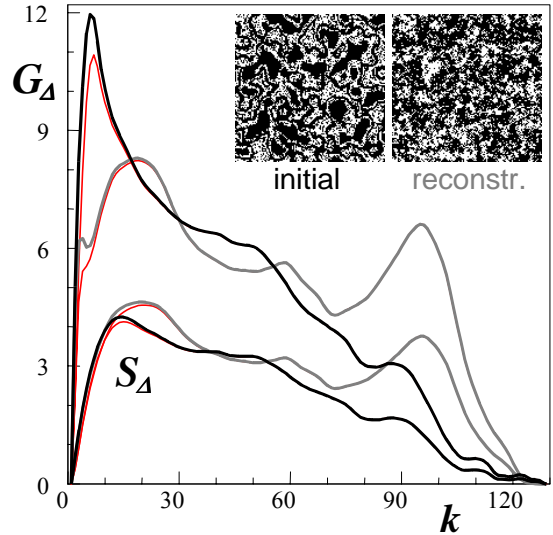

**Fig. 5.** The multiscale CoIN quantified by the entropic measure $G_\Delta(k)$ for the encoded in the shifted fashion (1-black, 256-white) initial laser speckle-pattern (upper black bold line) and reconstructed its counterpart (upper grey line), each of size $L \times L = 129 \times 129$ (in pixels) and 0.361 fraction of white phase (see the insets), adapted from Ref. [2] with the permission of the authors. When we use a kind of binary encoding (1-black, 0-white) then also the multiscale SpIN measure the $S_\Delta(k)$ can be used, correspondingly bottom (black and grey) lines. Easily seen discrepancies $\big| G_\Delta(k; \text{initial}) - G_\Delta(k; \text{reconstr.}) \big|$ and the similar ones for the $S_\Delta(k)$ substantiate the conclusion given in Ref. [2] about difficulty of obtaining of accurate reconstruction of the multiscale medium by using only the two-point correlation functions. For purpose of comparison, we display also the entropic descriptor $C_\lambda(k)$ [14] of a complex behaviour (thin lines). Notice, that the differences between $G_\Delta$ ($S_\Delta$) and $C_\lambda$ appear mostly at the initial length scales, which are structurally highly susceptible.

The author believe that with further development, this multiscale analysis could make the two-component entropic measure and the entropic descriptor of a complex behaviour [14] promising theoretical tools for various applications. For instance, they may provide information essential for further improvement of the reconstruction process [2] and its extension to greyscale patterns. Particularly interesting seems to be impact of gapped data in the context of the competition of two kinds of disorder, the standard spatial one and the novel compositional one. All this leads to the fresh 'multiscale perspectives' for systems of finite size objects and opens an interesting area for a future research.

## 4. Summary and conclusion

Into the model of pillars we incorporate only the basic aspects of the system of non-interacting finite size objects (with the exception of a kind of 'hard-core repulsion'):
a) The decomposability of the pillars built of unit cubes;
b) The conservation of the total number of pillars and total sum of their heights.

We have advanced two variants of a two-component entropic measure, based on the above model, $\tilde{S}_\Delta$ in Eq. (8) and $\tilde{S}_\Delta(h^*)$ in Eq. (11). The measure generalizes the entropic descriptors of grey-level inhomogeneity [13] and binary spatial inhomogeneity [6]. The current measure is devised for the multiscale analysis of the so-called spatio-compositional inhomogeneity (SpCoIN) for systems of finite size objects (FSOs). For a given 1D toy model with some parameters fixed, the possible categories of the macrostates have been illustrated on pictorial diagram in Fig. 1. Some of the corresponding reference macrostates as a function of total number of pillars have been presented in details in Tab. 1. The competition of two components, the $S_{(max)}$ connected with the spatial disorder and $G_{(max)}$ linked with the compositional disorder, which affect in a different degree the maximal two-component overall entropy $\tilde{S}_{max}$, is illustrated in Fig. 2.

In turn, Figs. 3, 4 and 5 refer to the compositional inhomogeneity (CoIN) that prevails in experimental data. Making use of both variants of the measure, the $G_\Delta$ and $G_\Delta(h^*)$, the expected distinction between two profiles of fractional Brownian motion (fBm) with slightly different Hurst exponents, $H_1 = 0.4$ and $H_2 = 0.5$, is detected in Fig. 3. In accord with the physical intuition, the lower Hurst exponent of a profile (thus the higher its fractal dimension), the greater CoIN is found by each of the measures, e.g., $G_\Delta(h^*; H_1) > G_\Delta(h^*; H_2)$. On the other hand, for the same Hurst exponents the two measures show a nice correlation. Therefore, instead of the $G_\Delta(h^*)$, one can use the measure $G_\Delta$ that is qualitatively correct but less demanding for 2D systems, in particular.

Fig. 4 deals with the initial 12-fold greyscale quasipattern evolving in time through the intermediate pattern of undefined symmetry to the final 14-fold greyscale quasipattern [28]. Here, the compositional inhomogeneity uncovers an approximate compositional greyscale periodicity, even for the intermediate pattern, which is absent, e.g., in strictly binary quasi-crystal planar structure, cf Fig. 4 in Ref. [7]. Due to this feature, the behaviour of the $G_\Delta$ for the three patterns is surprisingly regular along the length scales. But for the corresponding three curves that differ mainly in amplitudes the effect of multiple intersecting curves still appears. However, it should be stressed that in the CoIN the compositional degrees of freedom are involved exclusively while only the spatial ones are taken into account in the SpIN. So, when we are dealing with a greyscale quasipattern without any gaps we have the case (c) and, since there are no spatial dof, such the measures the $G_\Delta$ and $S_\Delta$ cannot be compared directly, in contrast to the case (b). This is possible also in the next example, where some kind of the greyscale-enhanced analysis is tested.

Finally, we test both measures, the $S_\Delta$ and $G_\Delta$, for the case of two-valued images i) a binary laser-speckle pattern of complex multiscale structure and ii) its reconstruction that is the best one within the high accuracy applied procedure using two-point correlation functions $S_2(r)$ of the initial and target medium [2]. In order to apply the $G_\Delta$ both images were further encoded in the shifted fashion (1-black, 256-white). In Fig. 5 the lack of the accurate reconstruction is clearly confirmed by the discrepancies in the values of the corresponding curves. In particular, the greyscale-enhanced procedure reveals the additional well marked the first peak in the $G_\Delta$ and also in $C_\lambda$. This justifies the suggestion that at the most interesting range of length scales the initial pattern is more complex than its reconstructed counterpart. As the origin of the $G_\Delta$ measure is slightly different from the $S_\Delta$ one, specific details can be detected by a one only of the two measures. This is an example of a complementary behaviour.

Given relevant examples illustrate the ability of the proposed entropic measure that is based on the simple model of decomposable pillars to quantify some of multiscale specific features occurring in complex patterns or more generally in systems of FSOs, which can be easily missed, are hardly to discern or invisible to other approaches.

## Appendix

For a $d$-dimensional system of pillars of fixed integer heights $1 \leq j \leq w$, when each of the pillars is treated as the whole entity, the most general macrostate describes a set $M(k) \equiv \{m_1(i, k), \ldots, m_j(i, k), \ldots, m_w(i, k)\}$, $i = 1, 2, \ldots, \kappa(k)$, where $w$ denotes the maximal allowable pillar height in a system. Here $m_j(i, k)$ is the multiplicity of pillar appearance of height $j$ at $i$th cell for a given length scale $k$. Thus, the $i$th cell occupation number by such pillars reads $n_i(k) = \Sigma_{j=1}^{w} m_j(i)$ and the corresponding sum of the pillar heights equals to $H_i(k) = \Sigma_{j=1}^{w} m_j(i) j$. The number of equally probable the appropriate microstates can be written as

$$\widetilde{\Omega}(k, M) = \prod_{i=1}^{\kappa} \left\{ \binom{k^d}{k^d - m_0(i, k)} \frac{\left[ \sum_{j=1}^{w} m_j(i, k) \right]!}{\prod_{j=1}^{w} [m_j(i, k)]!} \right\}, \tag{A.1}$$

where the number $m_0(i, k)$ of unoccupied positions at $i$th cell, the $k^d$ and $n_i(k)$ are connected to each other by the simple relation $k^d = m_0(i) + n_i(k)$. The combinatorial term in Eq. (A.1) accounts for possible positions of the $n_i(k)$ pillars while the second one refer to the number of compositions of the local sum $H_i(k)$ of pillar heights into exactly $n_i(k)$ positions with the specified multiplicities $m_j(i, k)$.

## Acknowledgements

The author would like to thank A. R. Rucklidge for providing the original image files of evolving quasipatterns presented in part in Fig. 4. I am also grateful to S. Torquato for sending the genuine text files for the binary laser-speckle patterns depicted in Fig. 5.